\documentclass{nic-series}

\newcommand{\aap}{A\&A}  
\newcommand{\aj}{AJ}      
\newcommand{\apjl}{ApJL}   
\newcommand{\apjs}{ApJS}   
\newcommand{\mnras}{MNRAS} 

\newcommand{\pasj}{PASJ}   
\newcommand{\pasp}{PASP}   
\newcommand{\araa}{ARA\&A} 

\newcommand{\hii}{H\,\textsc{ii}}
\newcommand{\msun}{\ensuremath{\mathrm{M}_{\odot}}}

\begin{document} 

\title{The Circumstellar Medium of Massive Stars in Motion}

\author{Jonathan Mackey \inst{1} \and Norbert Langer \inst{1} \and Dominique M.-A.\ Meyer \inst{1} \and Vasilii V.\ Gvaramadze \inst{2} \and Shazrene Mohamed \inst{3} \and Hilding R.\ Neilson \inst{4} \and Andrea Mignone \inst{5}}

\institute{Argelander-Institut f\"ur Astronomie, Auf dem H\"ugel 71, 53121 Bonn, Germany\\
         \email{\{jmackey, nlanger, dmeyer\}@astro.uni-bonn.de}
         \and
         Sternberg Astronomical Institute, Lomonosov Moscow State University, Universitetskij Pr.~13, Moscow 119992, Russia. \quad
         \email{vgvaram@mx.iki.rssi.ru}
         \and
         South African Astronomical Observatory, P.O.~Box 9, Observatory, Cape Town, 7935, South Africa. \quad
         \email{shazrene@saao.ac.za}
         \and
         East Tennessee State University, Box 70652, Johnson City, TN, 37614, USA.\\
         \email{neilsonh@etsu.edu}
         \and
         Universit\`a degli Studi di Torino, Via Pietro Giuria, 1, 10125 Torino, Italy.\\
         \email{mignone@ph.unito.it}
          }

\maketitle

\begin{abstracts}
The circumstellar medium around massive stars is strongly impacted by stellar winds, radiation, and explosions.
We use numerical simulations of these interactions to constrain the current properties and evolutionary history of various stars by comparison with observed circumstellar structures.
Two- and three-dimensional simulations of bow shocks around red supergiant stars have shown that Betelgeuse has probably only recently evolved from a blue supergiant to a red supergiant, and hence its bow shock is very young and has not yet reached a steady state.
We have also for the first time investigated the magnetohydrodynamics of the photoionised \hii{} region around the nearby runaway O star $\zeta$ Oph.
Finally, we have calculated a grid of models of bow shocks around main sequence and evolved massive stars that has general application to many observed bow shocks, and which forms the basis of future work to model the explosions of these stars into their pre-shaped circumstellar medium.
\end{abstracts}

\section{Introduction}
\label{sec:HBN23_Intro}
Massive stars are the main drivers of the evolution of the gaseous component of galaxies.
They emit huge quantities of far-ultraviolet photons that ionise and heat surrounding gas; their strong winds drive shocks and gas flows; and their final explosions as supernovae or gamma-ray bursts generate powerful blastwaves that eject chemically enriched matter into the interstellar medium (ISM).
They evolve from a hot main sequence star to a more extended red supergiant (RSG) or blue supergiant (BSG), and sometimes to more exotic Wolf--Rayet or Luminous Blue Variable phases\cite{Lan12}.
These changes induce strong variations in stellar wind properties, luminosity, and temperature, profoundly affecting the star's circumstellar medium (CSM) interaction and leading to a highly structured CSM by the time the star ends its life\cite{GarLanMac96}.

\begin{figure}[t]
\begin{center}
\includegraphics[width=0.63\textwidth]{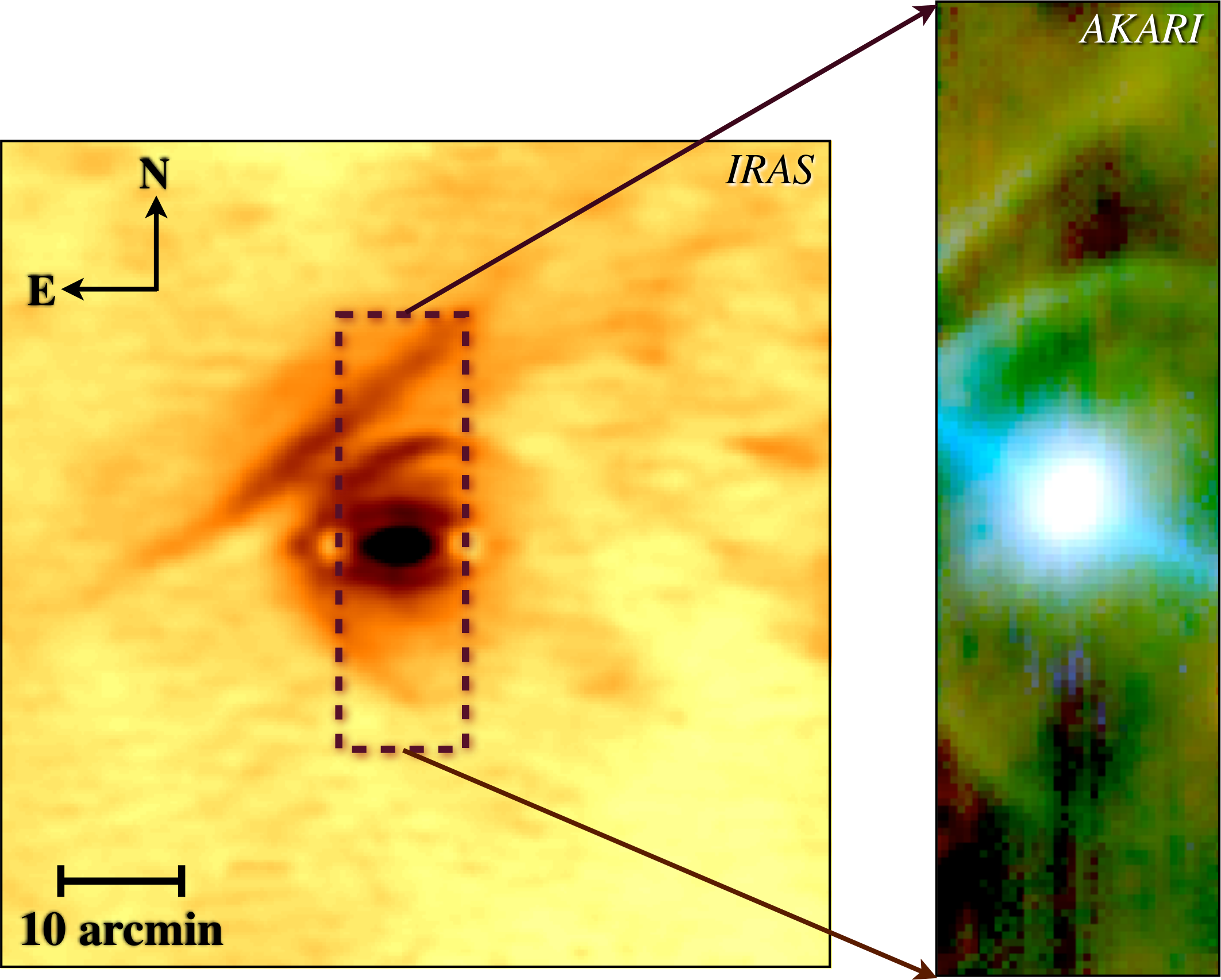}
\caption{\label{fig:HBN23_Betelgeuse}
  Far-infrared \textit{IRAS}\cite{CaoTerPri97} (60 $\mu$m) and \textit{AKARI}\cite{UetIzuYamEA08} (65 $\mu$m) images of the bow shock and linear bar in the circumstellar medium of Betelgeuse, taken from fig.~(1) of Mohamed et al.\cite{MohMacLan12}.
  Betelgeuse itself is the over-exposed central object; the elliptical feature surrounding it in the \textit{IRAS} image is an artefact.
  North is up, East is to the left, and the star is moving in a North-Easterly direction.
}
\end{center}
\end{figure}

Massive stars are born in groups and star clusters, but a large fraction of them are ejected from their birth sites either by dynamical interactions with other stars inside the cluster, or by the disruption of a binary system when one of its components explodes.
These stars, known as \emph{runaways} or \emph{exiles}, move through the ISM supersonically and this modifies the otherwise spherical CSM to a roughly axisymmetric solution with an upstream bow shock and a downstream wake\cite{Wil96}.
These structures have been observed for a number of runaway stars\cite{NorBurCaoEA97,GvaKroPfl10, PerBenBroEA12} in both our Galaxy and the Magellanic Clouds.
Examples are shown for the RSG Betelgeuse in Fig.~(\ref{fig:HBN23_Betelgeuse}), for the nearby main sequence star $\zeta$ Oph in Fig.~(\ref{fig:HBN23_ZetaOph}), and for the interaction of Supernova 1987A with its pre-shaped CSM in Fig.~(\ref{fig:HBN23_SN1987a}).

Our research group has worked since the 1990s to develop a comprehensive understanding of the interaction between massive stars and the various environments in which they form.
Our current project has focussed on how relative motion between the star and its surroundings affects the star--CSM--ISM evolution, investigating the differences between these models and previously calculated stationary star models.
We are also beginning to address the interaction of the star's supernova explosion with this structured CSM, to see what observational consequences this will have.
In this report we highlight the main results obtained in the past two years.

\section{The Bow Shock Around the Nearby Red Supergiant Betelgeuse}
\label{sec:HBN23_Betelgeuse}

\begin{figure}[t]
\begin{center}
\includegraphics[width=0.8\textwidth]{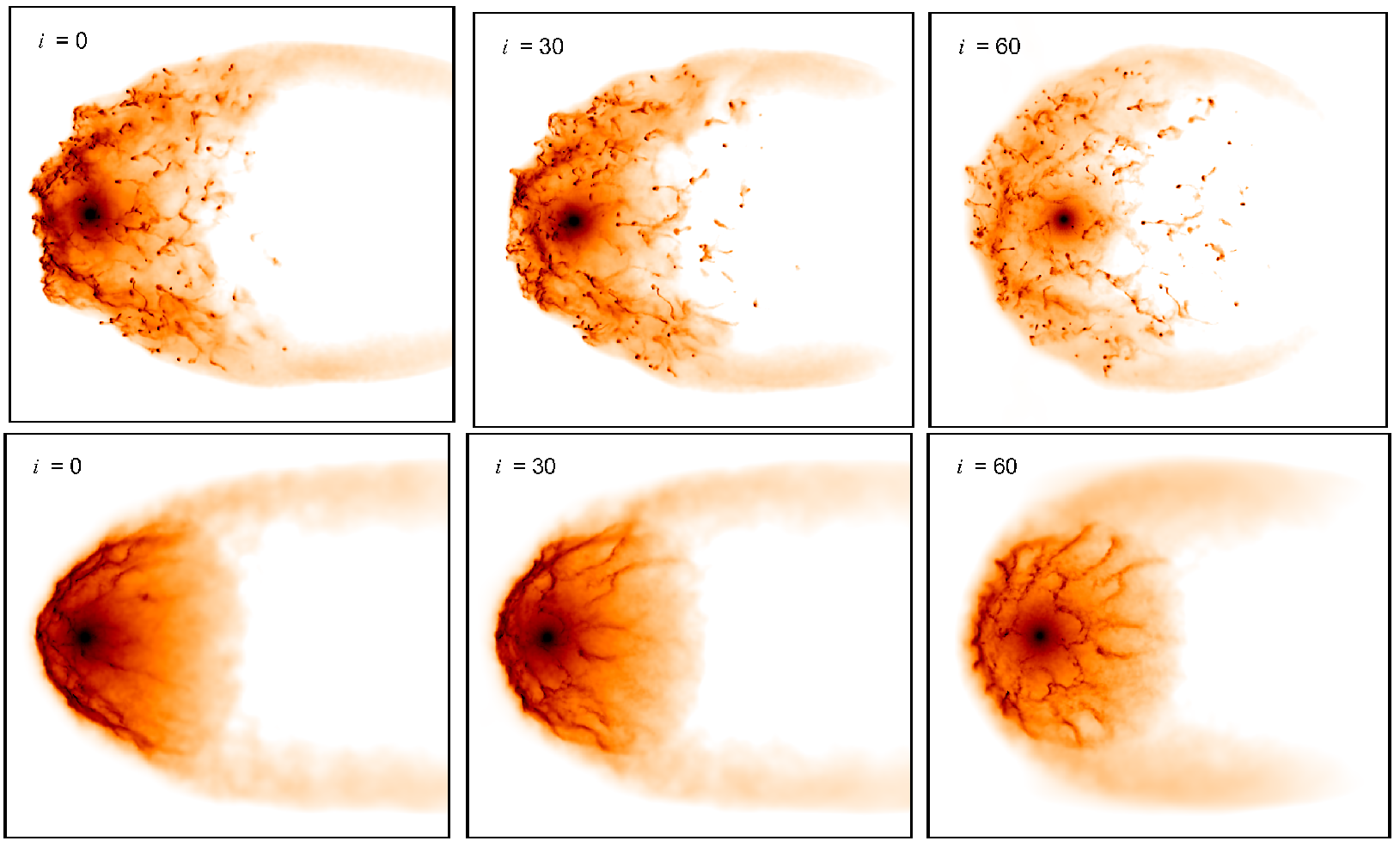}
\caption{\label{fig:HBN23_MML12}
  3D hydrodynamic simulations of RSG bow shocks (from fig.~12, Mohamed et al.\cite{MohMacLan12}) for a star moving with 32 km\,s$^{-1}$ (above) and 72 km\,s$^{-1}$ (below), viewed edge-on (left), at 30$^\circ$ inclination (centre), and at 60$^\circ$ inclination (right).
  Hydrogen column density is plotted on a logarithmic scale.
}
\end{center}
\end{figure}

Betelgeuse is one of the two nearest RSGs to Earth, and is the brightest star in the constellation Orion, ``The Hunter''.
It is also a well-known runaway star, and emits a powerful stellar wind that (because of the star's supersonic motion) generates a cometary shaped bow shock at its interface with the surrounding ISM\cite{NorBurCaoEA97} (see Fig.~\ref{fig:HBN23_Betelgeuse}).
Its CSM has a number of unexplained features that we aimed to investigate:
(1) The bow shock appears very smooth compared to numerical predictions\cite{UetIzuYamEA08,MohMacLan12,DecCoxRoyEA12};
(2) it is close to circular\cite{UetIzuYamEA08}, in contrast to predictions\cite{Wil96}; and
(3) there is a mysterious bar-shaped structure perpendicular to the star's direction of motion, at a larger radius than the bow shock\cite{NorBurCaoEA97}.

In Mohamed, Mackey, \& Langer\cite{MohMacLan12} we performed the first three-dimensional (3D) hydrodynamic simulations of the formation and evolution of Betelgeuse's bow shock, using a modified version of the hydrodynamics code \textsc{gadget-2}\cite{Spr05}.
We simulated RSGs moving at velocities ranging from 28 to 73 km\,s$^{-1}$ through a uniform ISM.
Results from two simulations are shown in Fig.~(\ref{fig:HBN23_MML12}), showing bow shocks at two different velocities from three viewing angles.

These ground-breaking simulations highlighted areas of significant agreement and disagreement with observations of Betelgeuse.
The bow shock appeared more smooth and layered for higher velocity stars, in closer agreement with observations.
We calculated the bow shock mass from \textit{AKARI} observations\cite{UetIzuYamEA08} and found a surprisingly low value ($\approx3\times10^{-3}\,\msun$) compared to the simulated bow shocks that are about 30 times more massive in steady state ($\approx0.1\,\msun$).
The shape of the simulated bow shock deforms from approximately circular at early times to the predicted cometary shape\cite{Wil96} when it reaches steady state.
Furthermore, the bow shock is smoother when it is young because instabilities have a finite growth time.
All of these findings support the hypothesis that Betelgeuse has entered the RSG phase only `recently' (within the last 25 thousand years), and that the young bow shock is still forming and expanding into its surroundings.
The low mass of the bow shock was subsequently confirmed by more detailed observations\cite{DecCoxRoyEA12}.

\section{Bow Shocks Around Moving Stars as They Evolve}
\label{sec:HBN23_BSG_RSG}

\begin{figure}[t]
\begin{center}
\includegraphics[width=1.0\textwidth]{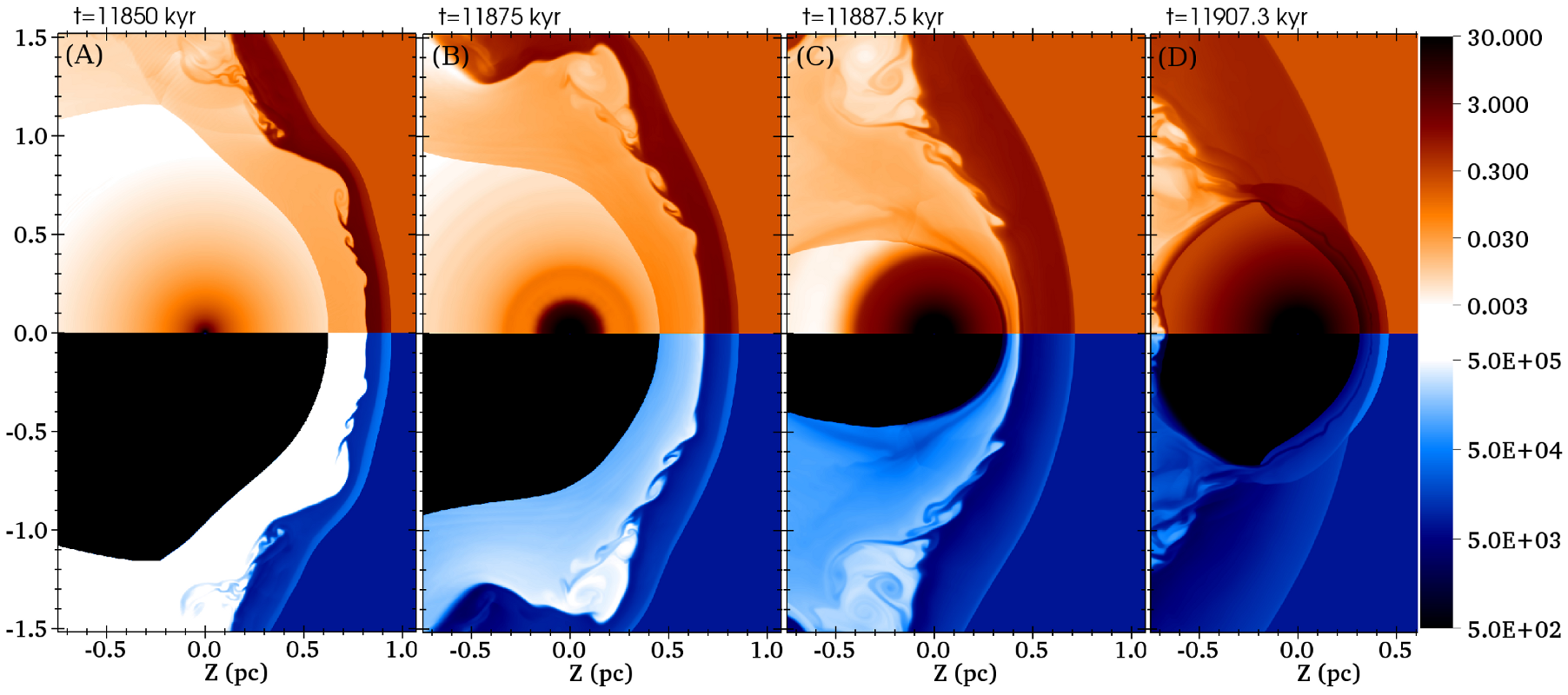}
\caption{\label{fig:HBN23_MMNLM12}
  2D simulation of double bow shocks as a BSG star (A) evolves to a RSG star (B and C) and to the end of its life (D).
  Gas number density $n_{\mathrm{H}}$ (upper half-plane, in $\mathrm{cm}^{-3}$) and temperature (lower half-plane, in Kelvin) are plotted on logarithmic scales at times $t=11.85,\ 11.875,\ 11.8875,$ and $11.9073$ Myr from left to right, respectively (reflecting the age of the star at each snapshot).
  The star is moving from left to right in each panel.
  This is a reproduction of fig.~(3) from Mackey et al.\cite{MacMohNeiEA12}.
}
\end{center}
\end{figure}

If Betelgeuse has only recently evolved from a BSG to a RSG, some remnants of the bow shock from this previous stage of evolution may still be imprinted in its CSM.
The linear bar outside the bow shock in Fig~(\ref{fig:HBN23_Betelgeuse}) is an obvious candidate, being oriented almost exactly perpendicular to Betelgeuse's direction of motion.
To test this possibility, we incorporated an evolving stellar wind module into the radiation-hydrodynamics code \textsc{pion}\cite{MacLim10}, and performed two-dimensional (2D) simulations of bow shocks around a moving star as it evolves from a blue to a RSG\cite{MacMohNeiEA12}.
The evolving wind properties were obtained directly from a stellar evolution model of a similar-mass star to Betelgeuse.

The results of one of the simulations are plotted in Fig.~(\ref{fig:HBN23_MMNLM12}).
Four snapshots are plotted corresponding to:
\textbf{(A)} the BSG phase, in which the hot star produces a fast and strong wind that supports a large bow shock;
\textbf{(B)} the beginning of the RSG phase where the star begins to emit a slow, dense and powerful wind;
\textbf{(C)} the interaction phase where the collapsing BSG bow shock generates a second, inner bow shock around the expanding RSG wind; and 
\textbf{(D)} the end of the star's life, by which time the two bow shocks have merged.

We identify panel (C) of Fig.~(\ref{fig:HBN23_MMNLM12}) with the current evolutionary phase of Betelgeuse and its CSM.
In this interpretation, the circular bow shock around Betelgeuse is the interaction of the newly-expanding dense RSG wind with the collapsing bubble of hot gas left by the (now extinguished) BSG wind.
The linear bar in Fig.~(\ref{fig:HBN23_Betelgeuse}) corresponds to the remnant BSG bow shock that is still upstream from the inner bow shock (but see also Ref.~\citen{DecCoxRoyEA12}).
The inner bow shock's mass, shape, thickness and size are consistent with the observed properties of Betelgeuse's bow shock, in contrast to steady-state models.

\begin{figure}[t]
\begin{center}
\includegraphics[height=0.35\textwidth]{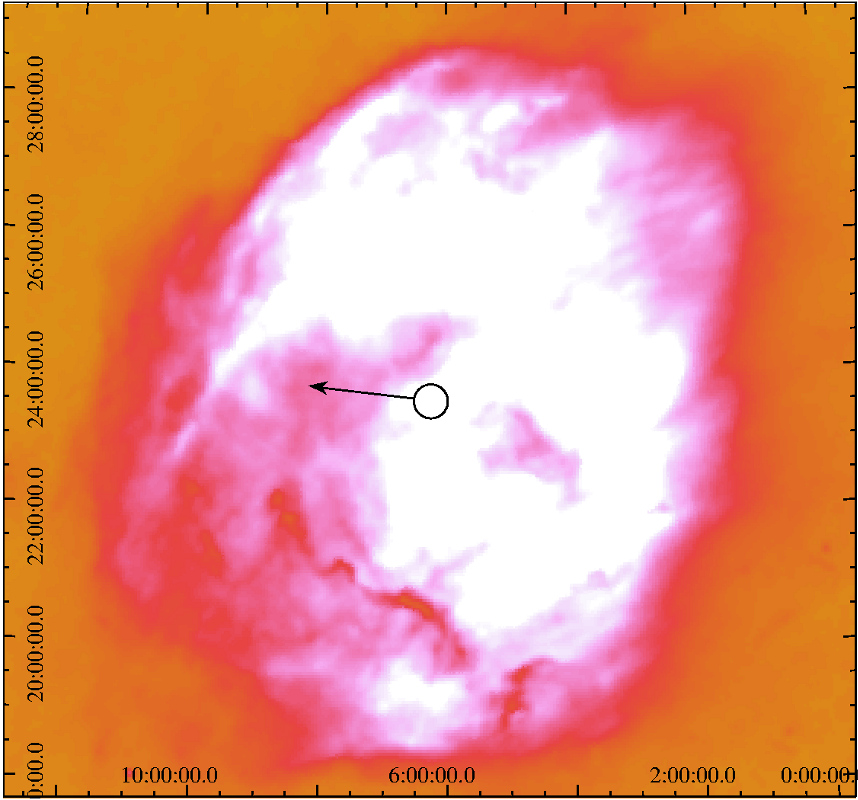}
\includegraphics[height=0.35\textwidth]{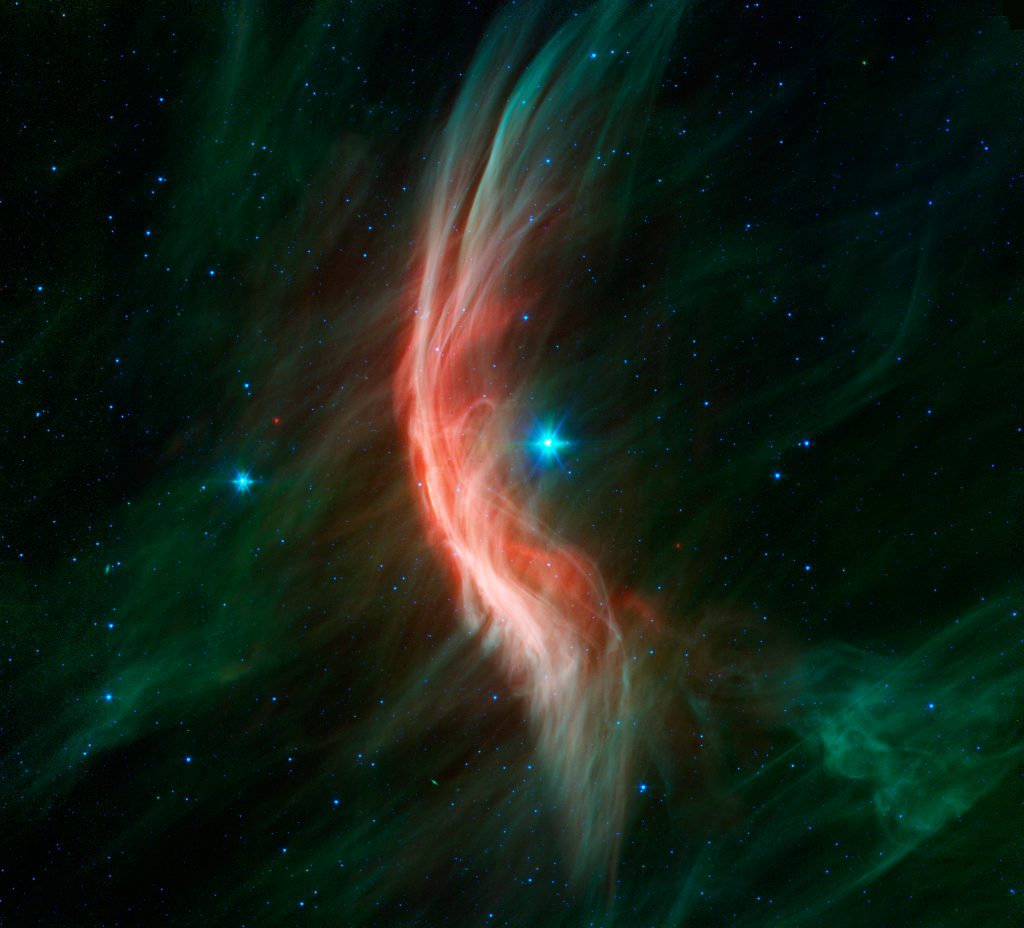}
\caption{\label{fig:HBN23_ZetaOph}
  Left: SHASSA H$\alpha$ image of the H\,{\sc ii} region around $\zeta$\,Oph, from fig.~(11) in Ref.~\citen{MacLanGva13}.
  The star is at the centre of the circle, and the arrow indicates its direction of motion relative to the ISM.
  The image is oriented with Galactic longitude (in units of degrees) increasing to the left and Galactic latitude increasing upwards.
  Right: \textit{Spitzer Space Telescope} image of the bow shock on a much smaller scale (the image spans about 1.5$^\circ$) at infrared wavelengths of $3.6-24\,\mu$m (Credit: NASA, JPL-Caltech, \textit{Spitzer Space Telescope}).
}
\end{center}
\end{figure}

\begin{figure}
\begin{center}
\includegraphics[width=0.4\textwidth]{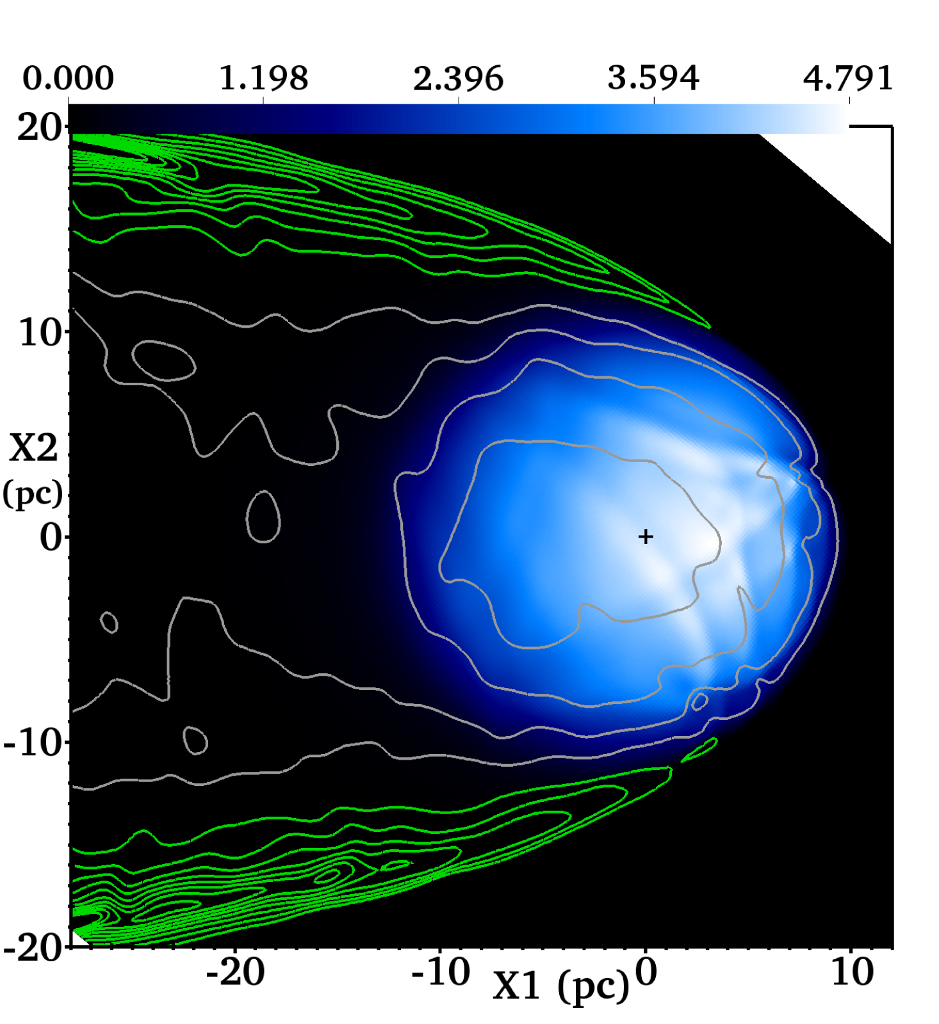}
\includegraphics[width=0.4\textwidth]{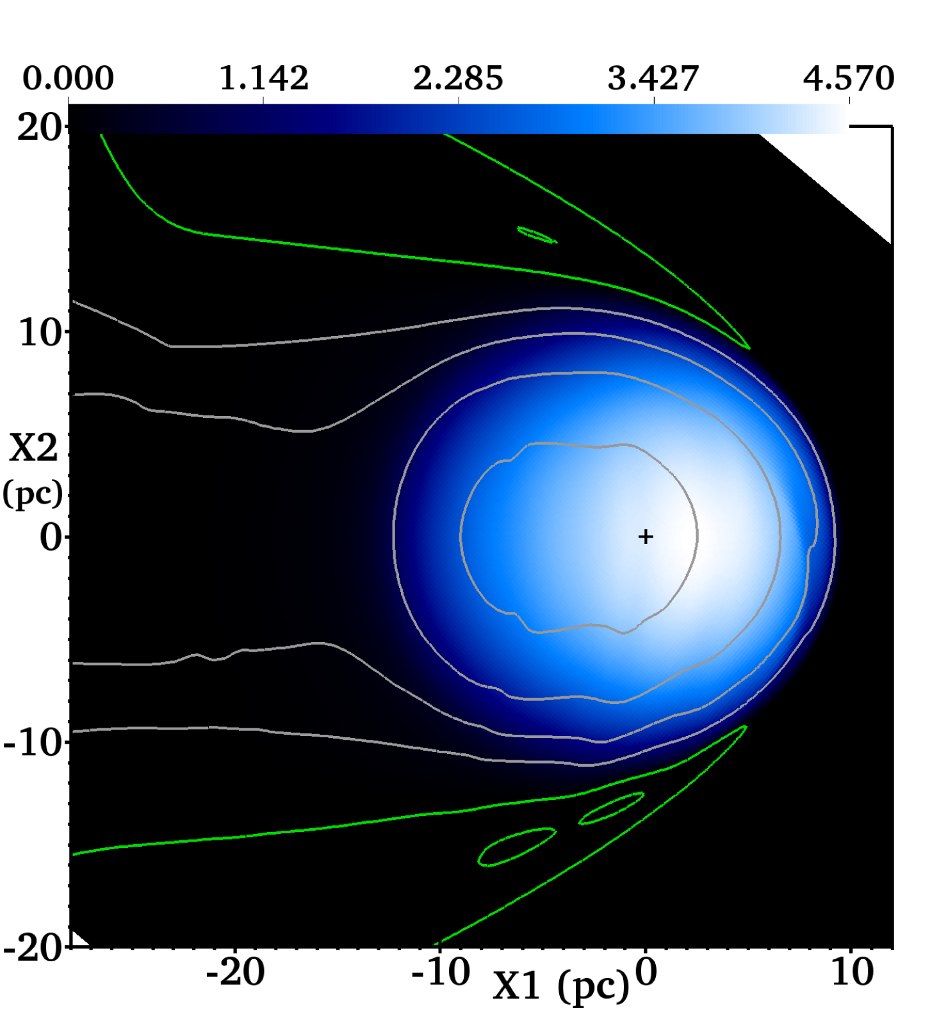}
\caption{\label{fig:HBN23_HII}
  Projections through the 3D simulations of \hii{} regions around a runaway O star for a non-magnetised ISM (left) and a magnetised ISM (right) where the magnetic field is along the line of sight (from figs.~4 and 6 of Ref.~\citen{MacLanGva13}).
  The star is marked with a cross, and is moving from left to right in the plane of the image.
  Projected H$\alpha$ emitted intensity is plotted on the linear colour scale and neutral H column density ($N_{\mathrm{HI}}$) as contours.
  The mean column density of the undisturbed grid is subtracted off (to remove grid edge effects), so underdense regions with negative column density are shown with grey contours and overdense regions with green contours.
}
\end{center}
\end{figure}

\section{\hii{} Regions Around Moving Main Sequence Stars}
\label{sec:HBN23_HIIregions}

The nearest example of an H\,\textsc{ii} region around an exiled star is Sh\,2-27, powered by the main sequence O star $\zeta$ Oph.
The left-hand panel of Fig.~(\ref{fig:HBN23_ZetaOph}) shows the H$\alpha$ image of this H\,\textsc{ii} region from the Southern H$\alpha$ Sky Survey Atlas (SHASSA)\cite{GauMcCRosEA01}.
The right-hand panel shows a spectacular infrared image of the bow shock around $\zeta$ Oph, on a much smaller scale.
In Gvaramadze, Langer \& Mackey\cite{GvaLanMac12} we showed that the relative sizes of the \hii{} region and bow shock could be used to constrain the mass-loss rate of $\zeta$ Oph.

In Mackey, Langer \& Gvaramadze\cite{MacLanGva13} we used the \textsc{pion} code to make the first 3D radiation-magnetohydrodynamics simulations of a star similar to $\zeta$ Oph moving through the ISM with a velocity of 26.5\,km\,s$^{-1}$.
Synthetic observations through the simulations are shown in Fig.~(\ref{fig:HBN23_HII}), where we plot the H$\alpha$ emission from ionised gas
(units: $10^{-16}\,\mathrm{erg}\,\mathrm{cm}^{-2}\,\mathrm{s}^{-1}\,\mathrm{arcsec}^{-2}$) and neutral hydrogen column density contours (with spacing $\Delta N_{\mathrm{HI}}=0.5\times10^{20}\,\ensuremath{\mathrm{cm}^{-2}}$).
We found that the \hii{} region generates a dense expanding shell expanding from its lateral surfaces, leaving a conical shell (with a large momentum) enclosing an underdense wake trailing behind it.
This shell should be observable in neutral hydrogen emission at 21 cm, although its properties are quite sensitive to the interstellar magnetic field.
The ionisation front in the upstream direction is also unstable, and may be more stable for the magnetised case than for the hydrodynamic case.

\section{Evolution and Explosion of Massive Stars in Their Environment}
\label{sec:HBN23_Grid}

\begin{figure}[t]
\begin{center}
\includegraphics[width=0.42\textwidth]{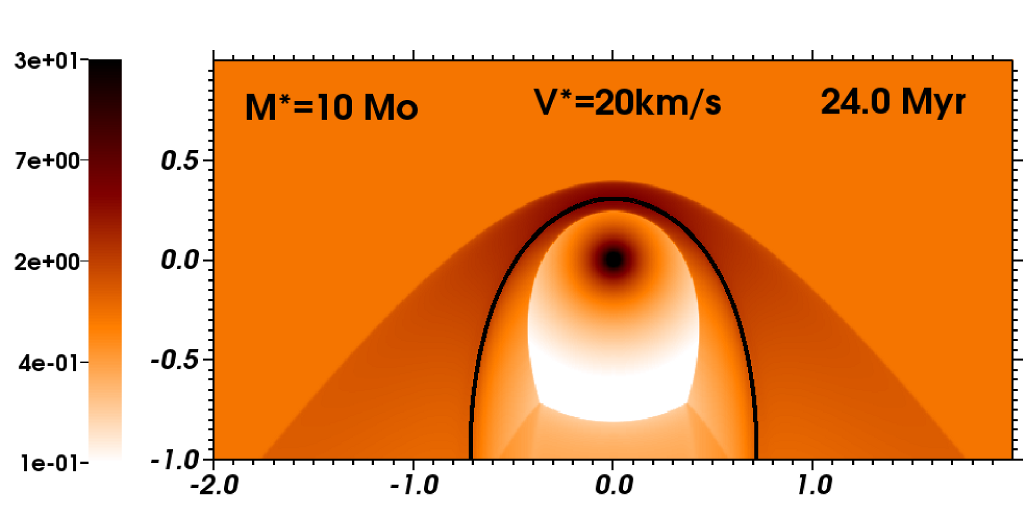}
\includegraphics[width=0.42\textwidth]{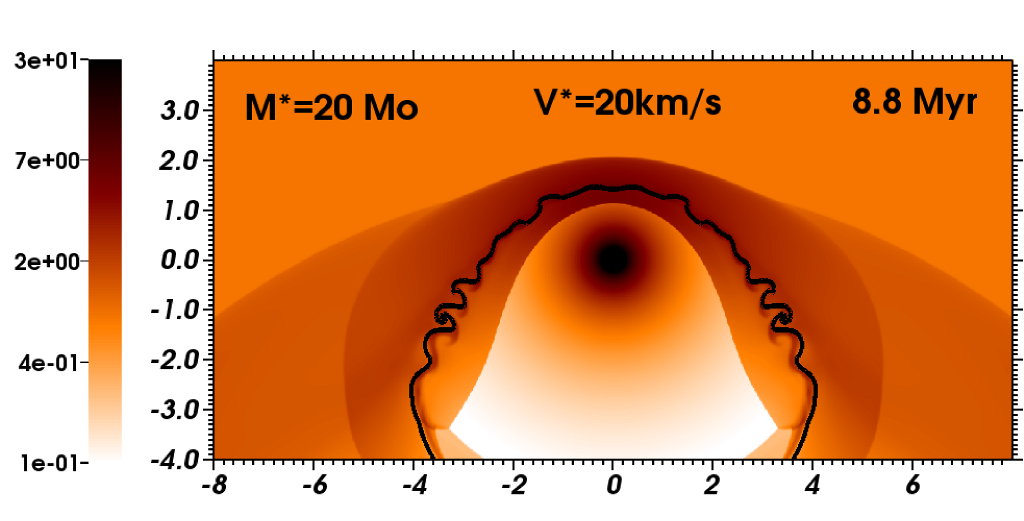}\\
\includegraphics[width=0.42\textwidth]{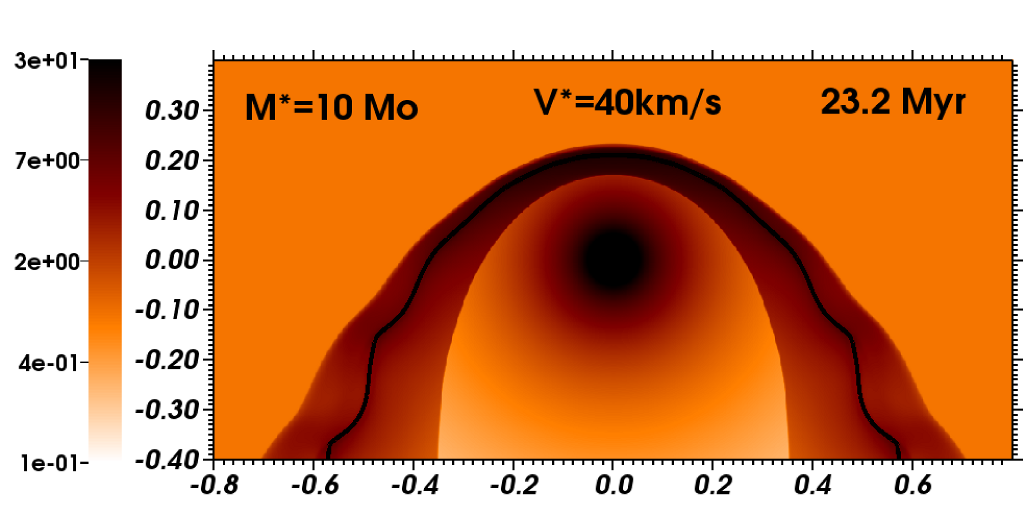}
\includegraphics[width=0.42\textwidth]{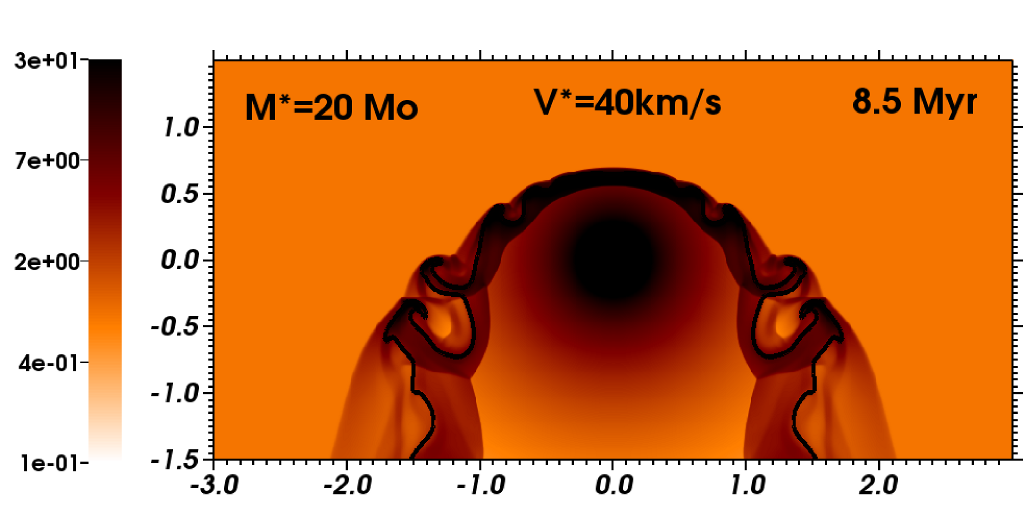}\\
\includegraphics[width=0.42\textwidth]{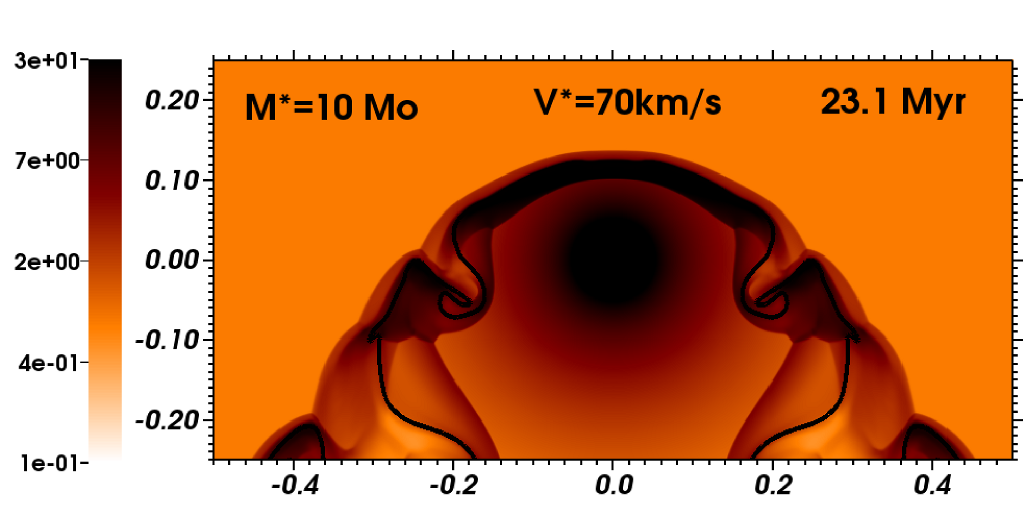}
\includegraphics[width=0.42\textwidth]{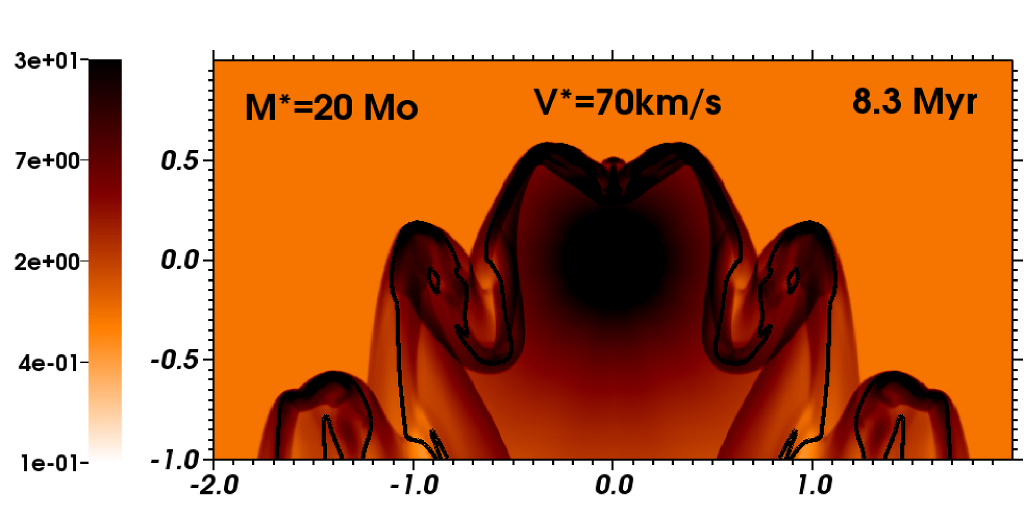}
\caption{
  Grid of models for stellar wind bow shocks from the RSG phase of a $10\, M_{\odot}$ (left panels) and $20\, M_{\odot}$ (right panels) initial mass stars according to their space velocity with respect to the ISM, with 20 (top panels), 40 (middle panels), and $70\, \mathrm{km}\, \mathrm{s}^{-1}$ (bottom panels).
  Keys denote the initial mass of the central star, the space velocity and the time from the zero age main-sequence, respectively.
  Gas number density is shown with a range from $0.1$ to $30.0\, {\rm cm}^{-3}$.
  Solid black contours trace the boundary between wind and ISM material.
  The x-axis is the radial direction and the y-axis the direction of stellar motion (both in pc).
  \label{fig:HBN23_GridRSG}
}
\end{center}
\end{figure}

We are using the magnetohydrodynamics code \textsc{pluto}\cite{MigZanTzeEA12} to simulate bow shocks around massive stars moving through the ISM for their full evolution.
Nine systems have been simulated, representing the complete stellar evolution of three stars (with masses 10, 20, and 40 $\msun$)
moving with three different velocities through the ISM.
We included thermal conduction by hot electrons as well as heating and cooling physics appropriate for the different stellar evolutionary phases.
A paper based on our results is now at an advanced stage of preparation (Meyer et al., 2013).
Fig.~\ref{fig:HBN23_GridRSG} shows six of these simulations for the RSG stage of evolution.
The gas density field is plotted for the 10 and 20 $\msun$ stars at three different velocities, increasing from top to bottom.
The stability of the bow shocks is broadly consistent with other works, and we are investigating the radiative emission from these bow shocks to compare to observations such as those of Betelgeuse.


\begin{figure}[t]
\begin{center}
\includegraphics[width=0.45\textwidth]{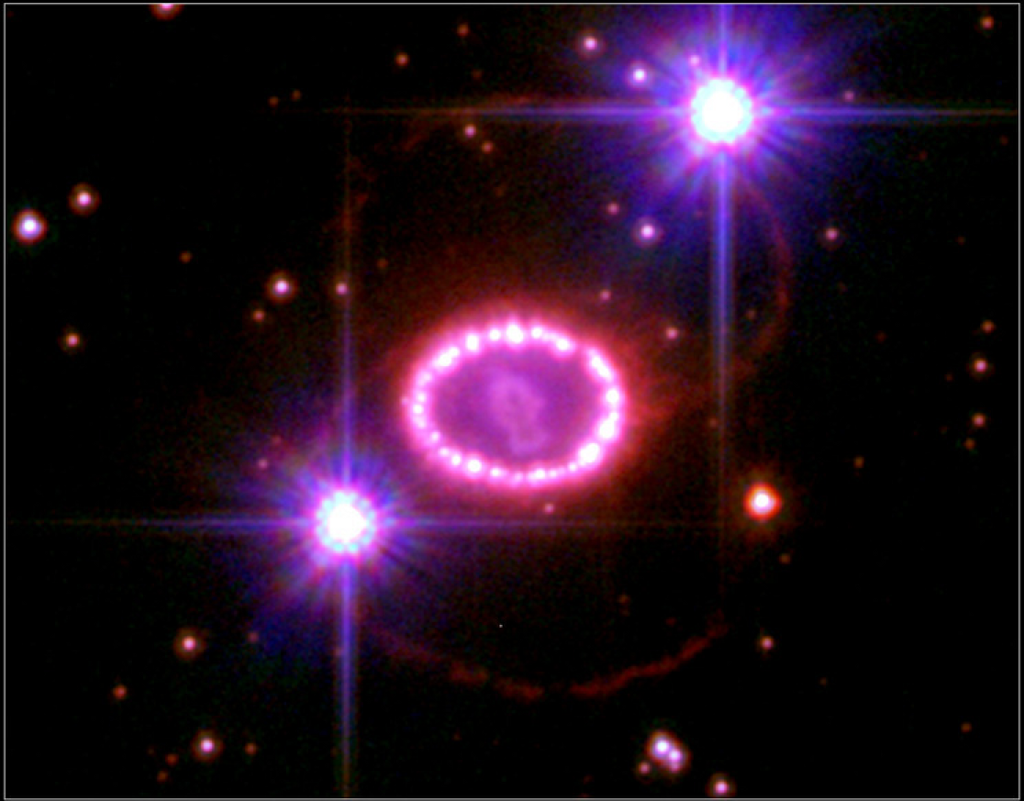}
\includegraphics[width=0.45\textwidth]{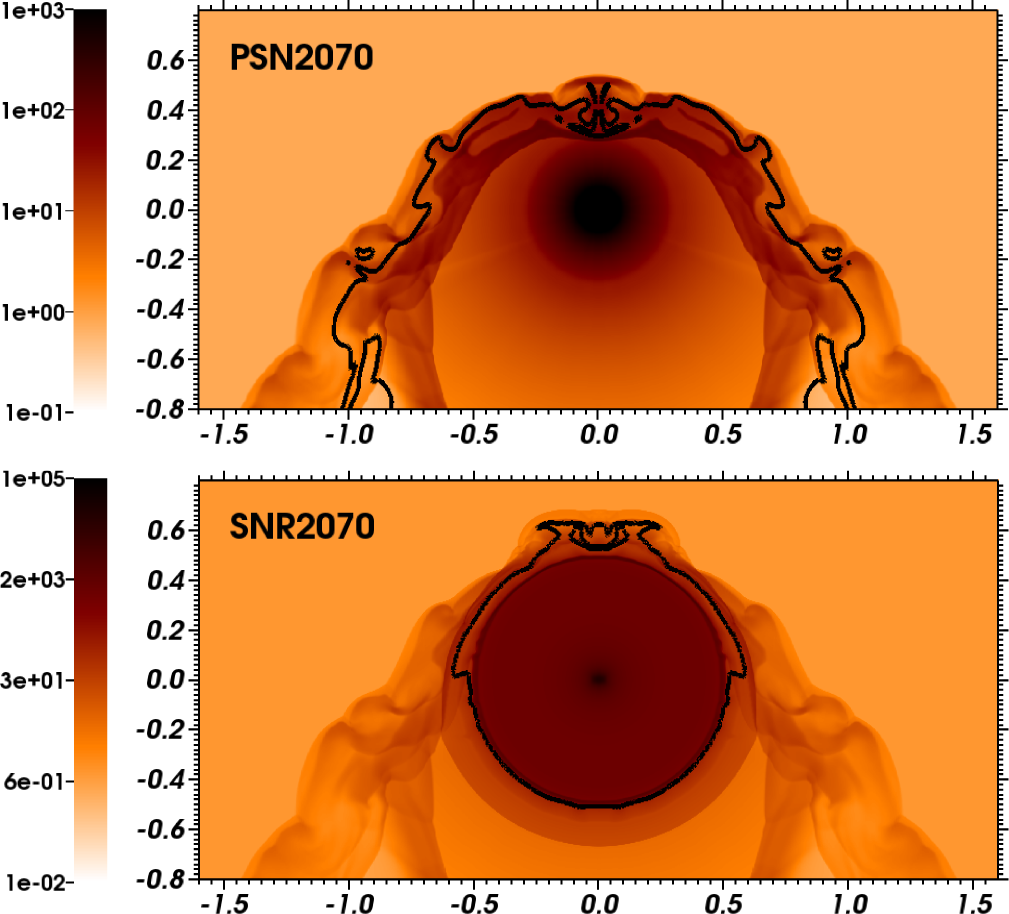}
\caption{Left: Supernova 1987A observed with the \textit{Hubble Space
  Telescope} (HST) in 2006.  (Credit: NASA, ESA, P.~Challis, R.~Kirshner, Harvard-Smithsonian Center for Astrophysics).
  Right: Simulation of the explosion of a supernova into the pre-shaped CSM around a runaway massive star. 
  The upper panel shows the pre-supernova density field of the gas (with hydrogen number density plotted on a logarithmic scale with units cm$^{-3}$), and a solid line showing the boundary between wind and ISM material.
  The lower panel shows a snapshot 18.7 years after the supernova explosion, when the blast wave from the star is ploughing through the bow shock (here the solid lines shows the boundary between supernova ejecta and CSM). 
  \label{fig:HBN23_SN1987a}
}
\end{center}
\end{figure}

We also calculate the CSM evolution up to the pre-supernova stage when the star is just about to explode.
This structured CSM then provides the initial conditions for our next project, to explode supernovae from the centre of the simulation and study the emission properties of the blast wave as it ploughs through the RSG wind and bow shock and on into the undisturbed ISM.
Following an established method\cite{vVeeLanVinEA09}, we first model the one-dimensional spherically-symmetric explosion of a supernova into a dense stellar wind. 
This is then mapped onto the centre of the 2D CSM grid and the blast wave expansion is followed to much larger radii.
An example of the pre-supernova bow shock and the subsequent supernova-CSM interaction is shown in Fig.~(\ref{fig:HBN23_SN1987a}), right-hand panel.
The blast wave is already distorted by the asymmetric CSM, and this should have observable consequences.

\section{Concluding Remarks}
\label{sec:HBN23_Conclusions}
The interaction of massive stars with their surroundings is an incredibly rich subject, where
a combination of the different stellar masses (with associated variations in evolution), different interstellar environments, and different stellar space velocities, produces a huge diversity of circumstellar structures.
Many of the closest such structures, such as the CSM of Betelgeuse and $\zeta$ Oph, have been observed in great detail with modern observatories, and yet we still do not have a definitive understanding of these objects.
The use of large supercomputers such as JUROPA has enabled us to make models with sufficient detail that we are now beginning to make quantitative comparisons to circumstellar structures, for both nearby stars and those in other galaxies.
This is significantly increasing our understanding of the physical processes involved and of the properties of the massive stars themselves, including some of the brightest and best known stars in the sky.

\section*{Acknowledgments}
We acknowledge the John von Neumann Institute for Computing for a grant of computing time on the JUROPA supercomputer at J\"ulich Supercomputing Centre.
JM and HRN acknowledge funding by fellowships from the Alexander von Humboldt Foundation.
Figs.~(\ref{fig:HBN23_Betelgeuse}) and (\ref{fig:HBN23_MML12}) reproduced with permission from Astronomy \& Astrophysics, \copyright{} ESO.



\end{document}